\documentclass[twocolumn,showpacs]{revtex4}

\begin{document}

  \newcommand{\Schlange}{\tilde}   

\title{
Gauge conditions for binary black hole puncture data \\
based on an approximate helical Killing vector
}

\author{Wolfgang Tichy, Bernd Br\"ugmann, Pablo Laguna}

\affiliation{
Center for Gravitational Physics and Geometry and 
Center for Gravitational Wave Physics\\
Penn State University, University Park, PA 16802}

\date{June 3, 2003}

\input epsf

\begin{abstract}
We show that puncture data for quasicircular binary black hole orbits allow
a special gauge choice that realizes some of the necessary conditions for
the existence of an approximate helical Killing vector field. Introducing
free parameters for the lapse at the punctures we can satisfy the condition
that the Komar and ADM mass agree at spatial infinity. Several other
conditions for an approximate Killing vector are then automatically
satisfied, and the 3-metric evolves on a timescale smaller than the orbital
timescale. The time derivative of the extrinsic curvature however remains
significant. Nevertheless, quasicircular puncture data are not as far from
possessing a helical Killing vector as one might have expected.
\end{abstract}

\pacs{04.20.Ex, 04.25.Dm, 04.30.Db, 04.70.Bw}
\maketitle

\section{Introduction}

Binary black hole (BH) mergers are among the most promising sources
for the ground-based interferometric gravitational wave detectors
GEO600, LIGO and TAMA~\cite{Schutz99}, which have started to collect
data. Predicting the gravitational waves from the dynamic and
non-linear regime of BH mergers requires numerical simulation, and a
crucial issue is to find astrophysically realistic initial data. This
is a non-trivial task which involves specification of some appropriate
free data and the solution of the constraint equations of general
relativity for the remaining data. By now several methods to produce
initial data for binary BH systems exist
(see~\cite{Cook:2000} for a recent review). However, there are still
open issues concerning the construction of astrophysically realistic
initial data, which in particular are data that represent two BHs on
almost circular orbits during their inspiral, and furthermore data
that represent a quantifiable approximation to two BHs near the
transition from inspiral to plunge and merger.

During the inspiral, we expect the two BHs to be in
quasicircular orbits around each other with a radius which shrinks on
a timescale much larger than the orbital timescale. This means that
the initial data should have an approximate helical Killing vector
$\xi$.  In addition one would like to have the initial data in
coordinates such that this approximate symmetry is manifest, i.e.\ the
time evolution vector should lie along $\xi$, so that the time
derivatives of the evolved quantities are minimized. Then we would get
\begin{equation}
\label{dtZeroApprox}
\partial_t g_{ij} 		\approx 
\partial_t \phi  		\approx 
\partial_t K 			\approx 
\partial_t {A}_{ij} 		\approx 0 ,
\end{equation}
where we have decomposed the metric into a conformal factor $\phi$ and
a conformal metric $g_{ij}$, and split the extrinsic curvature into
its trace $K$ and a tracefree piece ${A}_{ij}$. All the terms in
Eq.~(\ref{dtZeroApprox}) are of the order of some error quantity
which measures the failure of an exact helical Killing vector to
exist. This error has a finite value near the ``innermost stable
circular orbit'' and tends to zero for increasing separation of the
BHs. In principle one can improve the approximation by
replacing the circular orbits by inspiraling orbits defined
by a small but nonzero value of all the time derivatives
in Eq.~(\ref{dtZeroApprox}) based on post-Newtonian approximations,
see for example \cite{Tichy02}.

One approach to construct quasicircular initial data is to use the
conformal thin sandwich (CTS) decomposition~\cite{Wilson95,York99}. For
CTS data the time derivative of the conformal metric is free data that can
be set to zero. If in addition a maximal slicing lapse is used, one obtains
$
\partial_t g_{ij} 		= 
\partial_t K  			= 0
$.
Note, however, that $\partial_t \phi$ and $\partial_t {A}_{ij}$
are in general non-zero, since CTS by itself does not contain
conditions concerning helical Killing vectors. In the case of the
Meudon data~\cite{Gourgoulhon:2001ec,Grandclement:2001ed} 
(but see also~\cite{Pfeiffer:2002xz,Cook:2001wi}), CTS data is
constructed under the assumptions of a two-sheeted topology with
isometry, a conformally flat 3-metric, and a vanishing $K$.  In
addition, in order to construct quasicircular orbits, boundary
conditions on the lapse and shift are imposed such that the
Arnowitt-Deser-Misner (ADM) and Komar mass (computed from the lapse)
are equal, which is a necessary condition for the existence of a
helical Killing vector. The expectation is that these conditions yield
\begin{equation}
\label{dtZeroCTS}
\partial_t g_{ij} 		= 
\partial_t K  			= 0, \quad
\partial_t \phi			\approx 
\partial_t {A}_{ij} 		\approx 0.
\end{equation}
It is not clear a priori how well $\partial_t \phi \approx 0 $ and
$\partial_t {A}_{ij} \approx 0$ are satisfied since only one necessary
condition for the existence of the helical Killing vector has been
enforced, namely the equality of the ADM and Komar mass. 
It would be interesting to check how much $\partial_t \phi$
and $\partial_t {A}_{ij}$ deviate from zero. Compared to 
Eq.~(\ref{dtZeroApprox}), there are now additional errors in
Eq.~(\ref{dtZeroCTS}), which are related to the assumption
of conformal flatness, and to the details of
the construction. In particular the exact imposition of $\partial_t
g_{ij}=0$ and the method used for determining the orbital angular velocity 
of quasicircular orbits may introduce extra errors.

Another approach to construct binary BH initial data are puncture data
\cite{Brandt97b}, which are calculated using the conformal transverse
traceless (CTT) decomposition~\cite{York73,Cook:2000}. In this decomposition
$\partial_t g_{ij}=0$ cannot be achieved in general, while
$\partial_t K=0$ can always be imposed if a maximal slicing lapse is used.
There still remains the gauge
freedom to choose a shift that makes some of the time independence of a
helical approximate Killing vector manifest. So far approximate
helical Killing vectors have not explicitly been used in conjunction
with puncture data. Instead, in order to obtain quasicircular orbits
the effective potential method~\cite{Cook94} 
was used in the past.  In this method quasicircular
orbits are identified with the extrema in the binding energy. For thin
spherical shells of point particles it can be shown~\cite{Skoge02} that
both the effective potential method as well as equality of Komar mass
and ADM mass lead to the same identification of circular orbits. 
In general, however, it is not clear whether the effective potential
method leads to an approximate Killing vector in some sense.

The angular velocity $\Omega$ at the innermost stable circular orbit
(ISCO) computed from CTT puncture data with the effective potential
method is almost twice as large as for Meudon data. Since the $\Omega$
found in the Meudon data is close to post-Newtonian predictions, the
CTT method may be less reliable for ISCO data. It is important
to note, however, that if quasicircular CTT data is computed for the
separation predicted by the post-Newtonian and the Meudon CTS methods,
then the discrepancy in $\Omega$ amounts to only a few
percent~\cite{Baker:2002qf}. This is an indication that it is not the
CTT method by itself but rather the particular definition of the ISCO
that is problematic. The remaining discrepancy may have to do with the
different conformal decompositions used, but could also come from
different boundary conditions, or the different ways in which circular
orbits are found.  The key difference between quasicircular data in
the CTS and CTT methods is probably the construction of the extrinsic
curvature. For CTS it is chosen such that $\partial_t g_{ij}=0$, while
for CTT the Bowen-York extrinsic curvature is used with the hope that
the orbital parameters can be adjusted such that reasonable
approximations to circular orbits are obtained. An analytical
comparison of CTT data and CTS data can be found in~\cite{Laguna03}.

In this paper we focus on CTT puncture data, for which we want to
mention three reasons. First, the discrepancy between certain CTS and
CTT approaches is not fully understood, but CTT appears to be more
problematic. Second, in principle the puncture construction is
significantly simpler than current implementations of CTS with
excision. And third, to date all gravitational wave forms obtained
numerically for binary BH inspirals are based on puncture initial
data, see e.g.\ 
\cite{Alcubierre00b,Alcubierre02a,Baker:2001sf,Baker:2001nu,Baker:2003ds}.

In the following, we investigate the quasicircularity of CTT puncture
data numerically by asking whether such data possesses an approximate
Killing vector. In our analysis we do not make changes to the standard
CTT puncture data construction, but we try to find a lapse and shift
such that Eq.~(\ref{dtZeroApprox}) holds approximately. Concretely, we
first compute a particular puncture data set for a quasicircular orbit
as defined by the effective potential method. We then compute a
maximal slicing lapse with novel boundary conditions that are derived
from necessary conditions for the existence of a Killing vector,
namely the equality of the Komar and ADM masses at infinity or the
punctures.  Given the puncture data and the lapse, we calculate a
shift which minimizes $\partial_t g_{ij}$ in the sense that a certain
divergence of the metric does not evolve in time, obtaining
\begin{equation}
\label{dtZeroCTT}
\partial_t\partial_j g^{ij}     = 
\partial_t K  			= 0, \quad
\partial_t g_{ij} 		\approx
\partial_t \phi			\approx 
\partial_t {A}_{ij} 		\approx 0.
\end{equation}
In other words, we use the four degrees of freedom in lapse and shift
to set the time derivatives of four metric and curvature quantities to
zero. Compared to Eqs.~(\ref{dtZeroApprox}) and (\ref{dtZeroCTS}), in
addition to the assumptions about the Killing vector and conformal
flatness there will be non-vanishing terms due to the choice of
Bowen-York extrinsic curvature and the effective potential method. We
therefore check numerically how small the time derivatives are that
are supposed to vanish approximately in Eq.~(\ref{dtZeroCTT}), in
particular we use the Baumgarte-Shapiro-Shibata-Nakamura (BSSN)
system~\cite{Shibata95,Baumgarte99} of evolution equations to quantify
the magnitude of the time derivatives.

The paper is organized as follows. In Sec.~\ref{Puncture_CTT} we describe
puncture data and briefly explain the methods we use
to construct them numerically. Sec.~\ref{helical_KVs} describes
how the helical Killing vector assumption can be used to construct a lapse
and shift for puncture data and presents numerical results.
In Sec.~\ref{Discussion} we conclude with a discussion.

{\it Notation.}
In this paper we use the standard $3+1$ decomposition of Einstein's
equations, in which the 4-metric is written as
\begin{equation}                                               
ds^2 = -{\alpha}^2 dt^2 + \bar{g}_{ij}(dx^i +\beta^i dt)(dx^j +\beta^j dt).
\end{equation}
Here $\bar{g}_{ij}$ is the intrinsic 3-metric on a $t=const$ hypersurface 
and ${\alpha}$ and $\beta^i$ denote lapse and shift.
Spatial components of tensors are denoted by Latin indices from the middle
of the alphabet (e.g. $i$, $j$, ...). Latin letters from the beginning of
alphabet are used for spacetime tensor indices.
The extrinsic curvature is defined by 
\begin{equation}
\bar{K}_{ab} = -\frac{1}{2} {\pounds}_n \bar{g}_{ab} ,
\end{equation}
where $n^a$ is the unit normal to the $t=const$ hypersurface.
This implies that the time evolution vector is given by
\begin{equation}
\left( \frac{\partial}{\partial t} \right)^a = {\alpha} n^a + \beta^a .
\end{equation}
The 3-metric is decomposed into a conformal factor $\phi$ and
a conformal 3-metric $g_{ij}$, such that
\begin{equation}
\bar{g}_{ij} = \phi^4 g_{ij} .
\end{equation}
The extrinsic curvature $\bar{K}_{ij}$ is split into its trace $K$ and
its tracefree part $\bar{A}^{ij}$ by writing it as
\begin{equation}
\bar{K}^{ij} = \bar{A}^{ij} +\frac{1}{3} \bar{g}^{ij} K .
\end{equation}
As a rule, physical quantities have an overbar while conformal
quantities do not, but lapse and shift are denoted by $\alpha$ and
$\beta^i$ and are not rescaled.

\section{Puncture data and the CTT decomposition}
\label{Puncture_CTT}

Puncture data are computed using the 
conformal transverse traceless (CTT) decomposition.
In the CTT decomposition $\bar{A}^{ij}$ is decomposed into
\begin{equation}                                      
\bar{A}^{ij} = \phi^{-10} \left( A^{ij}_{TT} + LW^{ij} \right) ,
\end{equation}
where $A^{ij}_{TT}$ is an arbitrary transverse traceless
piece (i.e.\ $\nabla_j A^{ij}_{TT}=0$), and 
\begin{equation}
LW^{ij}= \nabla^i W^j + \nabla^j W^i -\frac{2}{3} g^{ij} \nabla_k W^k
\end{equation}
is the longitudinal piece.
Then the Hamiltonian and momentum constraints become
\begin{eqnarray}                                      
\label{ham_CTT}
  \nabla^2 \phi -\frac{1}{8} \phi R
  -\frac{1}{12}  \phi^5 K^2          &&	\nonumber \\
+\frac{1}{8} \phi^{-7}
  (A^{ij}_{TT} +LW^{ij} )
  (A^{kl}_{TT} + LW^{kl})g_{ik}g_{jl} 
&=&   0
\end{eqnarray}
and
\begin{equation}
\label{mom_CTT}
\nabla_{j} LW^{ij} -\frac{2}{3} \phi^6 \nabla^i K  = 0 ,
\end{equation}
which are elliptic equations for $\phi$ and $W^i$.

In the case of puncture data, the above CTS equations are 
further simplified by assuming
that the 3-metric is conformally flat, i.e.\
\begin{equation}
\bar{g}_{ij} = \phi^4 \delta_{ij} .
\end{equation}
In addition we set 
\begin{equation}
K = 0 
\end{equation}
so that the slice is maximal. We also assume that the transverse piece
vanishes, 
\begin{equation}
A^{ij}_{TT} = 0 . 
\end{equation}
Analytic solutions to Eq.~(\ref{mom_CTT}) 
are then given by
\begin{equation}  
W^{i} = \sum_{A=1}^2 \left[ 
        -\frac{1}{4r_A} \left( 7P^{i}_A + s^{i}_A s_{A j} P^{j}_A \right)
	\right]
\end{equation}    
and hence 
\begin{eqnarray}
\bar{K}^{ij} &=& \bar{A}^{ij} = \phi^{-10} LW^{ij}   \nonumber \\
             &=& \phi^{-10} \sum_{A=1}^2 \frac{3}{2r_A} \left(
2P^{(i}_A s^{j)}_A - (\delta^{ij} - s^{i}_A s^{j}_A)s_{Ak} P^k_A 
\right) \nonumber \\
\end{eqnarray}
fulfills the momentum constraint. 
The notation here is as follows:
The coordinate locations of the two  
BHs are denoted by $(x_1, y_1, z_1)$ and $(x_2, y_2, z_2)$ and we have
introduced   	  
\begin{equation}  
r_{A} = \sqrt{(x-x_{A})^2 + (y-y_{A})^2 + (z-z_{A})^2} .
\end{equation}    
and      	  
\begin{equation}  
s_{A}^i = (x-x_{A}, y-y_{A}, z-z_{A})/r_{A} ,
\end{equation}	  
where the subscript $A$ labels the BHs.
The parameters $P^i_A$ denote the momentum of each BH, and the
ADM momentum at infinity is $P^i_1+P^i_2$. 
On the other hand, the ADM momentum at the punctures, which represent
the inner asymptotically flat ends of the hypersurface, vanishes by
construction, i.e. the black holes `do not move' when viewed from the
inner ends which is a useful feature when looking for an approximate
Killing vector. Choosing $P^i_1+P^i_2=0$ we obtain data in coordinates
where the net linear momentum vanishes at all three spatial
infinities.

In order to find a conformal factor which fulfills the Hamiltonian
constraint we make the ansatz
\begin{equation}
\label{phiAnsatz} 
\phi  = 1 + \frac{m_1}{2r_1} + \frac{m_2}{2r_2} + u ,
\end{equation}
which together with Eq.~(\ref{ham_CTT}) leads to 
\begin{equation}
\nabla^2 u +\frac{1}{8} \phi^{-7} LW^{ij} LW^{kl} g_{ik}g_{jl}  =  0,
\end{equation}
and we impose 
\begin{equation}
\lim_{r\rightarrow \infty} u = 0
\end{equation}  	
as boundary condition on $u$. The equation for $u$ has to be solved
numerically. The solution $u$ is finite and at least twice
differentiable. It depends on the bare masses $m_A$, the momenta
$P_A$ and the separation $D=\sqrt{(x_1-x_2)^2 + (y_1-y_2)^2 +
(z_1-z_2)^2}$.  

In order to obtain quasicircular orbits, we use the parameters of
Tab.~\ref{parameters}.  The column denoted ISCO corresponds to the innermost
stable circular orbit (ISCO) as determined by Baumgarte~\cite{Baumgarte00a}
with numerical values as in~\cite{Baker00b}. The Pre-ISCO column is a stable
circular orbit found by Cook~\cite{Cook94} transcribed into puncture
data~\cite{Baker:2002qf}. Note that for ease of comparison we use the same
values for $D$, $|P_A|$, and $m_A$ as in
\cite{Baker00b,Baker:2002qf}, which were normalized such that the ADM
mass at infinity should be unity. However, we find that $M^{ADM}_{\infty}$
deviates from unity by roughly $1\%$. 
We thus estimate these parameters to be accurate to within about $1\%$,
meaning that according to the effective potential method our orbits are not
exactly circular. We were not able to obtain more accurate parameters from
the literature.

\begin{figure}
\epsfxsize=8.5cm 
\epsfbox{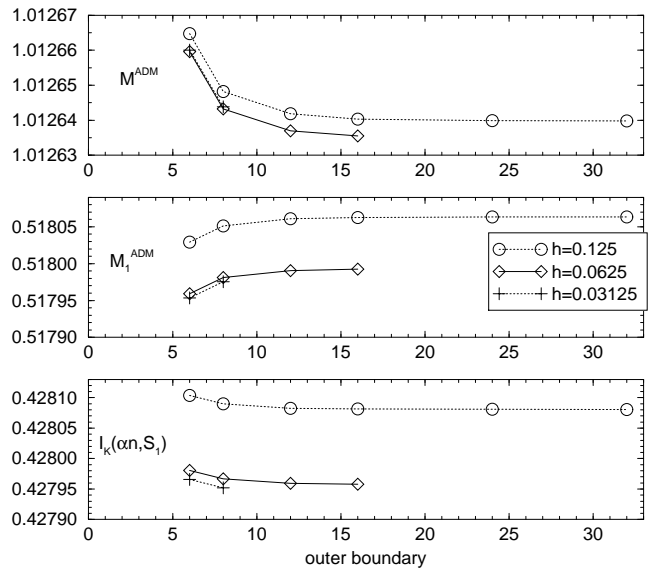}
\vspace{.4cm}
\caption{
Shown (from top to bottom) are the ADM mass at infinity, the ADM 
mass at one of the punctures and the Komar mass at one of the punctures,
each at three resolutions $h$ for different locations of
the outer boundary, in the case of the Pre-ISCO parameter set.
The error in the ADM mass at infinity at resolution $h=0.0625$ 
is about $0.0005\%$, while the ADM 
and Komar masses at the punctures have errors of $0.015\%$ and $0.030\%$,
}
\label{Mass_plots}
\end{figure}

Let us briefly discuss our numerical method.
We solve the elliptic equations numerically using second order finite
differencing and a multigrid elliptic solver. The code is a stand-alone
version of BAM\_Elliptic~\cite{Bruegmann99b}. All grids have uniform
resolution. Unless explicitly specified, all the results shown below
are obtained with the outer boundary located at $12$ and a 
finest resolution of $h=0.0625$.
As outer boundary conditions we use Robin conditions for
all scalars, i.e.\ we assume that $u \propto v
\propto 1/r$, where $r$ is the distance to the center of mass.  Here
$v$ determines the maximal slicing lapse (see Eq.~(\ref{alpha})
below). As boundary condition for the shift $\beta^i_0$ (computed below) we
use $\beta^i_0 \propto 1/r^2$, which is a simplifying assumption that works
reasonably well in practice.  Since we are dealing with punctures, no inner
boundaries are present. For the numerical work in this paper we consider
non-spinning equal mass binaries with their center of mass at rest at the
origin. The two BHs are positioned on the y-axis and their momenta point in
the positive and negative x-directions, resulting in an angular momentum
along the z-direction. After solving for $u$ and $v$ we observe second order
convergence to zero in the Hamiltonian constraint and in $\partial_t K$, as
expected. We also find second order convergence in the shift $\beta^i$ after
solving Eq.~(\ref{ellipt_beta}) below. The ADM and Komar 
integrals (defined in Sec.~\ref{helical_KVs} below) at
infinity are computed with volume integrals over $u$ and
$v$ covering the numerical domain plus a correction for the missing
volume such that the overall error is expected to fall off like one over the
distance to the outer boundary squared. At the punctures, ADM masses and
Komar integrals 
are obtained by fourth order interpolation of $u$ and $v$
onto the location of the punctures. From Fig.~\ref{Mass_plots} we see that
for $h=0.0625$ the error in our masses at infinity is about $0.0005\%$,
while the ADM and Komar integrals at the punctures have errors of $0.015\%$
and $0.030\%$ respectively.

\begin{table}
\caption{Parameters used in order to obtain quasicircular orbits 
within the effective potential method, with $\Omega_{eff}$ denoting
the inferred angular velocity.  The column denoted ISCO corresponds to
the innermost stable circular orbit (ISCO) as determined by
Baumgarte, and the Pre-ISCO column corresponds to a stable
circular orbit found by Cook. The $P_A=0$ column is Brill-Lindquist
data, i.e.\ two punctures with no momentum.
\label{parameters}}
\begin{tabular}{|l|l|l|l|}
\hline
Parameter set 		& ISCO			& Pre-ISCO 	& $P_A=0$\\
\hline
$D$ 			& 2.303			& 3.698 	& 3.698 \\
\hline
$\left| P_A \right|$ 	& 0.3350		& 0.2148 	& 0 \\
\hline
$m_A$ 			& 0.4500 		& 0.4775 	& 0.5 \\
\hline
$M^{ADM}_{\infty}$ 	& 1.003			& 1.013		& 1 \\
\hline
$J^{ADM}_{\infty}$	& 0.7715		& 0.7944	& 0 \\
\hline
$ M^{ADM}_{\infty} \Omega_{eff}$ & 0.176	& 0.102 	& 0 \\
\hline	
\end{tabular}
\end{table}

\section{Helical Killing vectors}
\label{helical_KVs}

After solving for $u$, the puncture data are completely determined
while lapse and shift are still arbitrary. We want to choose lapse
and shift such that there is as little dynamical evolution as
possible. This means we want a lapse and shift which satisfy
\begin{equation}
\label{dtZero}
\partial_t g_{ij} = \partial_t \phi = \partial_t K= \partial_t \bar{A}_{ij}
 =0
\end{equation}
as well as possible.
If Eq.~(\ref{dtZero}) really did hold,
it would imply the existence of a Killing vector
\begin{equation}
\label{hKV}
\xi^a = \left( \frac{\partial}{\partial t} \right)^a 
      = {\alpha} n^a + \beta^a ,
\end{equation}
which points along the time evolution vector.
Since we are considering orbiting binaries this Killing vector would have to
be a helical Killing vector, which means that at spatial infinity 
($r \rightarrow \infty$) 
\begin{equation}
\label{asymptKV}
n^a = T^a  \ \  \mbox{and} \ \ \beta^a=\Omega\Phi^a
\end{equation}
where $T^a$ and $\Phi^a$ are the asymptotic
time-translation and rotational Killing vectors at spatial infinity and 
$\Omega$ is the angular velocity with which the binary rotates.

\subsection{Choice of lapse}
\label{Lapse_choice}

In order to allow for the helical Killing vector (\ref{hKV})
we now proceed by choosing a lapse such that
\begin{equation}
\partial_t K =0 
\end{equation}
which leads to the elliptic equation
\begin{equation}
\label{maxLapse}
\nabla^2({\alpha}\phi) - {\alpha}\nabla^2\phi 
 = \phi^5 ({\alpha}\bar{K}_{ij}\bar{K}^{ij} + \beta^i \nabla_i K )
\end{equation}
for the lapse. Note that this equation is valid for any $g_{ij}$ and 
time independent $K$.
We now use the specific form of puncture data and make the ansatz
\begin{equation}		
\label{alphaphiAnsatz}  	
{\alpha}\phi = 1 - \left( \frac{c_1 m_1}{2r_1} +
         			\frac{c_2 m_2}{2r_2} \right) + v 
\end{equation}
for the lapse, where $v$ is a finite correction.
Together with Eqs.~(\ref{maxLapse}) and (\ref{ham_CTT}) this ansatz
yields the elliptic equation
\begin{equation}
\label{puncture_v}
\nabla^2 v = \frac{7}{8} 
({\alpha}\phi) \phi^{-8} LW_{ij} LW^{ij} ,
\end{equation}
which has to be solved numerically subject to the boundary condition 
\begin{equation}
\lim_{r\rightarrow \infty} v = 0 .
\end{equation}
Combining Eqs.~(\ref{phiAnsatz}) and (\ref{alphaphiAnsatz}) yields
\begin{equation}
\label{alpha}
{\alpha} =\frac{ 1 -  \left( \frac{c_1 m_1}{2r_1} + 
                                  \frac{c_2 m_2}{2r_2} \right) + v}
                   {1 + \frac{m_1}{2r_1} + \frac{m_2}{2r_2} + u} .
\end{equation}
Hence we obtain
\begin{equation}
\lim_{r \rightarrow \infty}{\alpha} = 1 
\end{equation}
as boundary condition at spatial infinity,
which is what one wants in asymptotically Minkowskian coordinates.
The value at each puncture is
\begin{equation}
\lim_{r_A \rightarrow 0}{\alpha} = -c_A,
\end{equation}
that is, our ansatz for the lapse (\ref{alphaphiAnsatz})
introduces the freedom to pick the value of the lapse at the inner
asymptotically flat ends. 

This freedom turns out to be essential when trying to satisfy the
mass condition for a helical Killing vector field. But let us first
discuss the situation for a single puncture without momentum, i.e.\
the case of $P^i_1=P^i_2=m_2=K_{ij}=u=0$, which corresponds to a
Schwarzschild BH in isotropic coordinates. Of course, the Schwarzschild
spacetime is static and thus has a Killing vector $\xi^a$ so that we
can ask for which $c_1$ Eqs.~(\ref{dtZero}) and (\ref{hKV}) do indeed
hold.  It is easy to see that
\begin{equation}
{\alpha} = 1 \ \ \mbox{everywhere} ,
\end{equation}
which corresponds to $v=0$ and $c_1=-1$, fulfills Eq.~(\ref{puncture_v}) so
that $\partial_t K =0$ holds. If in addition we choose $\beta^i=0$,
we also get $\partial_t g_{ij} = \partial_t \phi = 0$. However 
$\partial_t \bar{A}_{ij} \neq 0 $, and hence $c_1=-1$ is not a good choice
since it leads to an evolving $\bar{A}_{ij}$ 
even in the case of the static Schwarzschild spacetime.

Another possibility obtained from $c_1=1$ and $v=0$ is
\begin{equation}
\label{isotropicLapse}
{\alpha} =\frac{ 1 - \frac{m_1}{2r_1} }
                   { 1 + \frac{m_1}{2r_1} } ,
\end{equation}
which also yields $\partial_t K =0$. This is the standard lapse of isotropic
coordinates. If we again choose $\beta^i=0$ we find that
Eqs.~(\ref{dtZero}) and (\ref{hKV}) are now satisfied. Therefore $c_1=1$ is
a good choice for a single BH.
It turns out that for $c_1=1$ the Komar mass
$I_K({\alpha}n,S_{\infty})$ (see Eq.~(\ref{Komar}) below)
computed from ${\alpha}$
equals the ADM mass of the BH. This is not true for $c_1=-1$ for which
$I_K({\alpha}n,S_{\infty})=0$. 

Let us return to the general case of binary BH data and discuss
what it means to impose equality of the Komar and ADM masses. The binary
puncture data that we are considering may not possess a helical
Killing vector $\xi^a$, but if they did then the Komar integral
\cite{Komar59,Ashtekar79a,Wald84}
\begin{equation}
\label{Komar}
I_K(\xi,S) = -\frac{1}{8\pi} \int_S \bar{\epsilon}_{abcd} \bar{\nabla}^c \xi^d
\end{equation}
integrated over any closed 3-surface $S$ containing the punctures
would yield the same
answer. Furthermore if we choose $S$ to be a sphere 
at $r \rightarrow \infty$ and if $\xi^a$ is normalized such that
\begin{equation}
\label{xi_norm}
\lim_{r \rightarrow \infty} \xi^a n_a = -1 ,
\end{equation}
then
\begin{equation}
\label{Komar_ADM}
I_K(\xi,S) = M^{ADM}_{\infty} - 2\Omega J^{ADM}_{\infty} 
\end{equation}
is expected to hold as an immediate generalization of~\cite{Ashtekar79a}.
Here $M^{ADM}_{\infty}$ and $J^{ADM}_{\infty}$ are the ADM mass and angular
momentum at spatial infinity ($r \rightarrow \infty$) given by
\begin{equation}
\label{M_ADM}
M^{ADM}_{\infty} = P^{ADM}_{\mu \ \infty} P^{ADM}_{\nu \ \infty} 
\eta^{\mu\nu} ,
\end{equation}
\begin{equation}
\label{E_ADM}
P^{ADM}_{0 \ \infty} = \frac{1}{16\pi}\int_{S_{\infty}}
                    (\bar{g}_{ij,i}-\bar{g}_{ii,j} )d\bar{S}^j ,
\end{equation}
\begin{equation}
\label{P_ADM}
P^{ADM}_{i \ \infty} = \frac{1}{8\pi}\int_{S_{\infty}}    
                 (\bar{K}_{ij} - K\eta_{ij} )d\bar{S}^j ,                    
\end{equation}
and
\begin{equation}
\label{J_ADM}
J^{ADM}_{\infty}= \frac{1}{8\pi}\int_{S_{\infty}} 
               (\bar{K}_{ij}\Phi^j - K \Phi_i)d\bar{S}^i .
\end{equation}
Note that the normalization condition (\ref{xi_norm}) requires that
${\alpha} \rightarrow 1$ for $r \rightarrow \infty$.

If we insert $\xi^a$ as given in Eq.~(\ref{hKV}) into
Eq.~(\ref{Komar}) we obtain
\begin{eqnarray}
\label{Komar2}
I_K(\xi,S) &=& I_K({\alpha}n,S)+I_K(\beta,S)  \nonumber \\ 
&=& \frac{1}{4\pi}\int_{S}\bar{\nabla}_i{\alpha} d\bar{S}^i
   -\frac{1}{4\pi}\int_{S} \bar{K}_{ij} \beta^i d\bar{S}^j .
\end{eqnarray}
Using Eq.~(\ref{asymptKV}) we find that the second term 
integrated at $r \rightarrow \infty$ is
\begin{equation}
\label{Komarbeta_infty}
I_K(\beta,S_{\infty})= -2\Omega \frac{1}{8\pi}\int_{S_{\infty}}
                          \bar{K}_{ij}\Phi^j d\bar{S}^i
          = -2\Omega J^{ADM}_{\infty} .
\end{equation}
Therefore combining 
Eqs.~(\ref{Komar_ADM}), (\ref{Komar2}) and (\ref{Komarbeta_infty}) 
the condition
\begin{equation}
\label{MK_MADM}
I_K({\alpha}n,S_{\infty}) = M^{ADM}_{\infty} 
\end{equation}
must hold if the helical Killing vector of Eq.~(\ref{hKV})
exists.

We can now address the question what the values $-c_A$ of the lapse
at the punctures should be. Since we are interested in the case
where a helical Killing vector exists (at least approximately), we should
pick $c_A$ such that Eq.~(\ref{MK_MADM}) is fulfilled, otherwise
$\xi^a$ cannot be a Killing vector. 
Since Eq.~(\ref{MK_MADM}) is just one condition on in principle two
unknowns, we also set $c_1=c_2$ so that the lapse has the same value at each 
puncture. This is justified by the fact that in our numerical computations 
we only study equal mass binaries with $m_1=m_2$.
Of course Eq.~(\ref{MK_MADM}) is only a
necessary condition for a Killing vector. 
To test whether a Killing vector really exists, we have to check
how well Eq.~(\ref{dtZero}) is fulfilled.

We summarize our numerical results in Tab.~\ref{masses_Omega}. Given
the ISCO and Pre-ISCO parameters described in Tab.~\ref{parameters},
we iterate $c_A$ until $I_K({\alpha}n,S_{\infty}) =
M^{ADM}_{\infty}$. Tab.~\ref{masses_Omega} gives the values for $c_A$ for
which this can indeed be achieved. Since the lapse has now been fixed, we
proceed to check two other relations, namely the mass equality at the
punctures (as opposed to infinity) and the angular velocity $\Omega$
predicted by the Komar integral.

Note that Eq.~(\ref{Komar_ADM}) can be used to compute
the angular velocity $\Omega$ if the Komar integral 
$I_K(\xi,S) $ is evaluated on a surface $S\neq S_{\infty}$.
Let us use Eq.~(\ref{Komar2}) to calculate $I_K(\xi,S_p) $
for $S_p = S_1 \cup S_2$, 
where $S_1$ and $S_2$ are infinitesimally small spheres around
each puncture. If we set $\beta^i$ to zero 
at the punctures (an assumption which will be justified below in 
Sec.~\ref{Shift_choice}), 
the shift term in Eq.~(\ref{Komar2}) does not contribute
and we find
\begin{eqnarray}
\label{Komar_punc}
I_K(\xi,S_p) &=& I_K({\alpha}n,S_1) + I_K({\alpha}n,S_2) \nonumber \\
&=&    m_1 \left[\frac{1+c_1}{2} + \frac{ v_1 +c_1 u_1 }{2} 
				 + \frac{ (c_1 - c_2)m_2 }{4D}
\right] \nonumber \\
&& \mbox{} \!\! + m_2 \! \left[\frac{1+c_2}{2} + \frac{ v_2 +c_2 u_2 }{2}
				 + \frac{ (c_2 - c_1)m_1 }{4D} 
\right] \!\! , \nonumber \\  
\end{eqnarray}
where $u_A$ and $v_A$ are the values of $u$ and $v$ at puncture $A$.
Note that $c_1 = c_2$ implies that there is no explicit dependence on
$D$, but $u_A$ and $v_A$ depend on $D$.
If $\xi^a$ is a Killing vector, $I_K(\xi,S_p)$ should 
have the same value as in Eq.~(\ref{Komar_ADM}), which in turn 
implies
\begin{equation}
\label{OmegaFromI_K}
\Omega = \frac{M^{ADM}_{\infty} - I_K(\xi,S_p)}{2J^{ADM}_{\infty}} .
\end{equation}   
Comparing Tabs.\ \ref{parameters} and \ref{masses_Omega} we see that
the angular velocity (\ref{OmegaFromI_K}) of the binary is
very close to what is found with the effective potential method.

\begin{table}
\caption{The value of the lapse at the puncture 
for the three parameter sets of table \ref{parameters} is chosen such that
Eq.~(\ref{MK_MADM}) holds. We find that then Eq.~(\ref{MK_MADM1}) also
approximately holds. In addition the $\Omega$ 
of Eq.~(\ref{OmegaFromI_K}) is close to
what the effective potential method predicts, and up to numerical accuracy
$\Omega = \Omega_{\beta}$ for $\Omega_{\beta}$ introduced in 
Sec.~\ref{Shift_choice}.
\label{masses_Omega}}
\begin{tabular}{|l|l|l|l|}
\hline
Parameter set 			& ISCO	& Pre-ISCO & $P_A=0$ \\
\hline
$c_A$ 				& 0.726		& 0.829 & 1.000 \\
\hline
$M^{ADM}_{\infty}$ 		& 1.003		& 1.013 & 1.000 \\
\hline
$I_K({\alpha}n,S_{\infty})$	& 1.003		& 1.013 & 1.000 \\
\hline
$M^{ADM}_{A}$ 			& 0.514		& 0.518 & 0.534 \\
\hline
$I_K({\alpha}n,S_A)$	& 0.372		& 0.428 & 0.500 \\
\hline
$c_A M^{ADM}_{A}$			& 0.373		& 0.430	& 0.534 \\
\hline
$c_A M^{ADM}_{A} - I_K({\alpha}n,S_A) \over c_A M^{ADM}_{A} $
				& 0.0027	& 0.0047 & 0.0637 \\
\hline
$ M^{ADM}_{\infty} \Omega$ 	& 0.168		& 0.100 & 0.000\\
\hline
$ M^{ADM}_{\infty} \Omega_{\beta}$	& 0.168	& 0.100	& 0.000\\	
\hline
\end{tabular}
\end{table}

A further check is provided by the analogue  
\begin{equation}
\label{MK_MADM1}
I_K({\alpha}n,S_{A}) = c_A M^{ADM}_{A} 
\end{equation}
to Eq.~(\ref{MK_MADM}).
Here the ADM mass of each puncture is given by
\begin{equation}
M^{ADM}_{A} = m_A (1+ u_A) +  \frac{m_1 m_2}{2D}  ,
\end{equation}
and the factor $c_A$ comes from the fact that at the puncture $\xi^a$
is normalized such that $\xi^a n_a = c_A$, which differs from the
standard normalization in Eq.~(\ref{xi_norm}).  As one can see in
Tab.\ \ref{masses_Omega} our numerical results for $M^{ADM}_{A}$ and
$I_K({\alpha}n,S_{A})$ approximately satisfy Eq.~(\ref{MK_MADM1})
in the case of quasicircular orbits, while for the $P_A=0$ data
Eq.~(\ref{MK_MADM1}) is violated. Yet even for quasicircular orbits
the deviation from Eq.~(\ref{MK_MADM1}) is larger than our numerical
errors, but the parameters of Tab.~\ref{parameters} used to put the
punctures into quasicircular orbits (according to the effective
potential method) are not very accurate. We believe that these
parameters are accurate up to $1\%$ error, hence some deviations from
Eq.~(\ref{MK_MADM1}) are expected. In future work we plan to construct
quasicircular orbits within our numerical method, which involves
varying $P^i_A$, and it should be possible to define $P^i_A$ for
quasicircular orbits by imposing Eq.~(\ref{MK_MADM1}). For the
parameters used in the present work, we conclude that the one
parameter freedom of the lapse at the puncture allows us to find a
lapse that is approximately compatible with the two necessary
conditions (\ref{MK_MADM}) and (\ref{MK_MADM1}) for a helical Killing
vector.

\begin{figure}
\epsfxsize=8.5cm 
\epsfbox{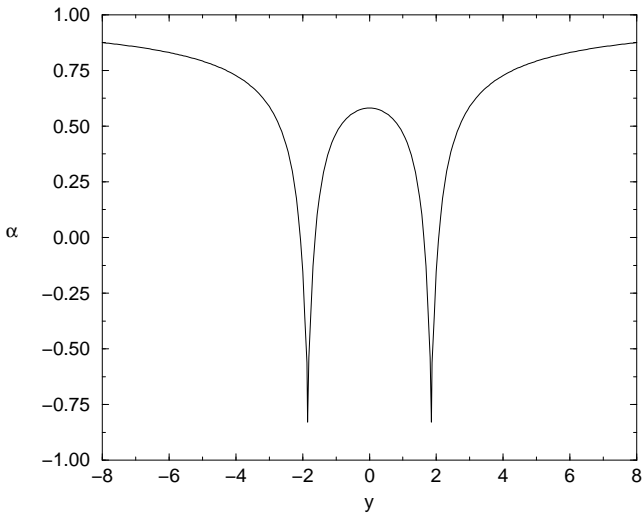}
\vspace{.4cm}
\caption{
The lapse ${\alpha}$ along the $y$-axis for the Pre-ISCO
data set. The punctures are on the $y$-axis at $y=\pm 1.849$.
}
\label{alphagraph}
\end{figure}

A plot of the lapse is shown in Fig.~\ref{alphagraph} for two punctures
located on the $y$-axis for the case of 
the Pre-ISCO parameter set of Tab.\ \ref{parameters}.

\subsection{Choice of shift}
\label{Shift_choice}
 
In Sec.~\ref{Lapse_choice} we have chosen the lapse such that
$\partial_t K=0$. Similarly we would like to choose a shift
which results in $\partial_t g_{ij} =0$. In general 
(if the conformal factor is chosen such that
$\partial_t \mbox{det} g_{ij} = 0$), we have
\begin{equation}
\label{g_evo}
\partial_t g_{ij} = - 2{\alpha}\phi^{-4}\bar{A}_{ij} + L\beta_{ij}  .
\end{equation}
Note however that $L\beta_{ij}$ is by definition purely longitudinal,
so that the left hand side can only be zero if   
$2{\alpha}\phi^{-4}\bar{A}_{ij}$ is purely longitudinal as well.
The CTS construction enforces this feature by exchanging the role of
$\bar{A}_{ij}$ and $\partial_t g_{ij}$, i.e.\ the time derivative of the
metric is treated as free data that defines the extrinsic curvature
through
\begin{equation}
\label{A_CTS}
\bar{A}_{ij} = \frac{1}{2{\alpha}\phi^{-4}} 
               \left( L\beta_{ij} - \partial_t g_{ij} \right).
\end{equation}
For maximal slicing and vanishing time derivative of the metric one
obtains an elliptic equation for the shift by taking a divergence of
(\ref{A_CTS}) and using the momentum constraint which leads to
\begin{equation}
\label{beta_CTS}
\nabla_j L\beta^{ij} =  L\beta^{ij} \nabla_j \log 
                           \left({\alpha}\phi^{-6} \right).
\end{equation}
In the CTS construction these are three of five coupled equations for
$\phi$, ${\alpha}$, and $\beta^i$.

However, for puncture data with maximal slicing, $\phi$,
${\alpha}$, and $\bar{A}_{ij}$ are determined independently of the
shift, and in particular $2{\alpha}\phi^{-4}\bar{A}_{ij}$ will not
be purely longitudinal in general. In fact, for the puncture
data we are considering here,
\begin{equation}
\label{apLW}
2{\alpha}\phi^{-4}\bar{A}_{ij} = 2{\alpha}\phi^{-6} LW_{ij} ,
\end{equation}
so that $2{\alpha}\phi^{-4}\bar{A}_{ij}$ is the product of a longitudinal
piece and a scalar function. 

But we can still determine a shift that completely removes the
longitudinal piece from the time derivative of the metric. 
Concretely, in order to find the longitudinal piece $L\beta^{ij}$
of Eq.~(\ref{apLW}) for puncture data, let us make the ansatz 
\begin{equation}
\label{apA_decomp}
2{\alpha}\phi^{-6} LW^{ij} = L\beta^{ij} + v_{TT}^{ij} ,
\end{equation}
where $v_{TT}^{ij}$ is a possible transverse traceless piece 
with $\nabla_j v_{TT}^{ij} =0$.
By taking the divergence of Eq.~(\ref{apA_decomp}) we find
the elliptic equation
\begin{equation}
\label{ellipt_beta}
\nabla_j L\beta^{ij} = \nabla_j \left( 2{\alpha}\phi^{-6} LW^{ij}\right) ,
\end{equation}
which has a unique solution for $\beta^{i}$ for given 
boundary conditions (compare with Eq.~(\ref{beta_CTS}) for CTS).
Once we have found $\beta^{i}$, we can then determine
$v_{TT}^{ij}$ from Eq.~(\ref{apA_decomp}).
Since we are interested in a binary configuration in corotating coordinates
we should adopt 
\begin{equation}
\label{betaboundary}
\lim_{r\rightarrow \infty} \beta^{i} 
= \Omega_{\beta} \Phi^i
\end{equation}
as boundary conditions for $\beta^{i}$ at spatial infinity. Here $\Phi^i$
is the asymptotic rotational Killing vector pointing along the direction of
rotation and $\Omega_{\beta}$ is the angular velocity of the binary.
Let us split the shift into the two pieces
\begin{equation}
\label{betasplit}
\beta^{i} = \beta^{i}_0 + \beta^{i}_{rot} ,
\end{equation}
where we have introduced the rotational piece
\begin{equation}
\label{betarot}
\beta^{i}_{rot}= \epsilon^{i}_{jk} x^j \Omega_{\beta}^k
\end{equation}
and a piece with 
\begin{equation}
\label{zeroboundary}
\lim_{r\rightarrow \infty} \beta^{i}_0 = 0 .
\end{equation}
The benefit of this split is that $L\beta_{rot}^{ij} = 0$, so that 
if $\beta^{i}_0$ fulfills Eq.~(\ref{ellipt_beta}) with 
boundary condition (\ref{zeroboundary}),
$\beta^{i}$ of Eq.~(\ref{betasplit})
immediately fulfills Eq.~(\ref{ellipt_beta})
with boundary condition (\ref{betaboundary}).
We have numerically solved for $\beta^{i}_0$ and then added 
a rotational piece of the form (\ref{betarot}) such that 
$\beta^{i}=0$ at each puncture. This is a very desirable property
since punctures have no linear momentum when viewed from the
asymptotically flat region at the puncture, so that the
natural choice for the shift is indeed zero.
\begin{figure}
\epsfxsize=8.5cm 
\epsfbox{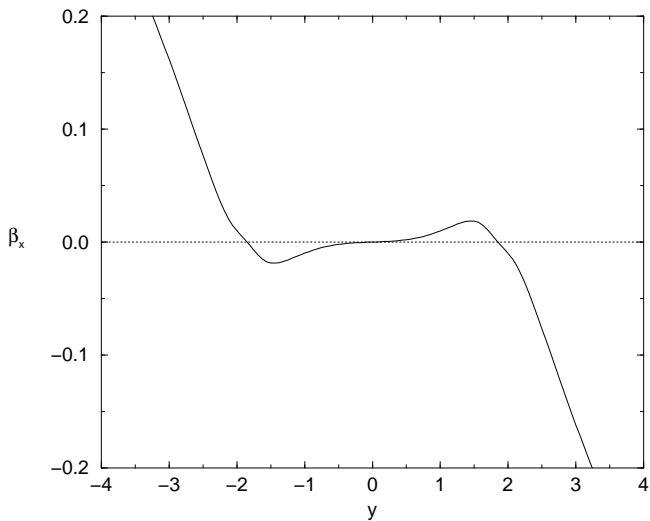}
\vspace{.4cm}
\caption{The shift $\beta^{x}$ for Pre-ISCO data along the $y$-axis, 
which connects the two punctures. At each puncture $\beta^{x}$ vanishes.  }
\label{betagraph}
\end{figure}
The shift 
for the Pre-ISCO parameter set of Tab.\ \ref{parameters} is shown in
Fig.~\ref{betagraph}. 
Notice that $\beta^{i}$ is linear for large $r$ in agreement with
boundary condition (\ref{betaboundary}). 
It turns out that up to the accuracy of our numerical computation,
the $\Omega_{\beta}$
required to achieve $\beta^{i}=0$ at each puncture is equal
to the $\Omega$ computed from Eq.~(\ref{OmegaFromI_K}).
We want to stress here that $\Omega_{\beta} = \Omega$ is another
necessary condition for a Killing vector 
(see appendix \ref{I_K_withouKV}) and not
expected to hold a priori for puncture data.

\begin{figure}
\epsfxsize=8.5cm
\epsfbox{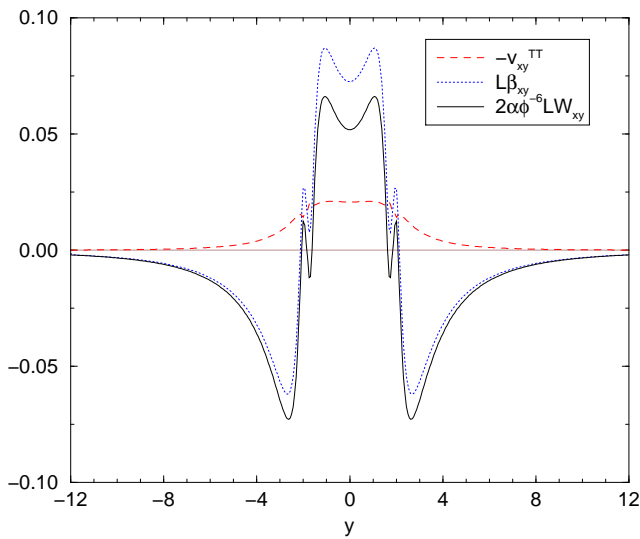}
\vspace{.4cm}
\caption{The three terms in Eq.~(\ref{apA_decomp}) for Pre-ISCO data.
}
\label{apA_decomp_graph}
\end{figure}
In Fig.~\ref{apA_decomp_graph} we show plots of the three terms in
Eq.~(\ref{apA_decomp}). We find that $v_{TT}^{ij}$ in
Eq.~(\ref{apA_decomp}) is indeed non-zero. 
Hence Eq.~(\ref{g_evo}) becomes
\begin{equation}
\partial_t g_{ij} = -2{\alpha}\phi^{-6} LW_{ij}  + L\beta_{ij} 
                  = - v^{TT}_{ij}  ,
\end{equation}
and $\partial_t g_{ij}=0$ cannot be achieved for puncture data.
Note however that $v_{TT}^{ij}$ is smaller than $L\beta^{ij}$, and
that $2{\alpha}\phi^{-6} LW^{ij}$ is purely longitudinal 
for large $r$. Thus $\partial_t g_{ij}$ is at least reduced by the above
choice of $\beta^i$. The maximum value is
\begin{equation}
\max \partial_t g_{ij} =  2\times 10^{-2} / M^{ADM}  .
\end{equation}
Hence the smallest timescale on which $g_{ij}$ evolves is
$T_{g_{ij}} = 2\pi (1 / \max \partial_t g_{ij}) \sim  300 M^{ADM} $
which is longer than the dynamical timescale given by
$T_{dyn} = 2\pi / \Omega \sim 60 M^{ADM} $. 
This means that $g_{ij}$ is approximately constant
up to an error of at most $T_{dyn} / T_{g_{ij}} \sim 20\% $ in the center
and much less away from the punctures.

So far we have only addressed the time derivatives of $K$ and $g_{ij}$.
We will now use the BSSN evolution equations to investigate
the remaining time derivatives. Recall that BSSN introduce the
additional variables
\begin{equation}
\Schlange{\Gamma}^i = - \partial_j \Schlange{\gamma}^{ij} ,
\end{equation}
where the BSSN metric $\Schlange{\gamma}_{ij}=g_{ij}=\delta_{ij}$ 
on the initial slice
for the case of conformally flat puncture data. Using
Eqs.~(\ref{g_evo}) and (\ref{apLW}) we obtain
\begin{equation}
\partial_t \Schlange{\Gamma}^i = -\nabla_j \partial_t g^{ij} 
= \nabla_j L\beta^{ij} - \nabla_j \left( 2{\alpha}\phi^{-6}LW^{ij}\right).
\end{equation}
Thus, when Eq.~(\ref{ellipt_beta}) is fulfilled we have
\begin{equation}
\label{Gammafreezing}
\partial_t \Schlange{\Gamma}^i = 0,
\end{equation}
so that $\Schlange{\Gamma}^i$ does not evolve with our choice of
shift. Put differently, we have three free functions available in
terms of the shift vector. This does not suffice to set $\partial_t
g_{ij} = 0$ for puncture data, but we can impose the so-called
Gamma-freezing condition (\ref{Gammafreezing})
\cite{Alcubierre00a,Alcubierre02a}, which we have obtained here
by constructing a shift for CTT data motivated by CTS data.

The time derivatives of the remaining variables have to be studied
numerically. We will do this next for the Pre-ISCO parameter set
of Tab.\ \ref{parameters}. First we look at the variable
\begin{equation}	
\varphi=\log \phi ,
\end{equation}  
which BSSN introduce by decomposing the physical metric as 
$\bar{g}_{ij} = e^{4\varphi} \Schlange{\gamma}_{ij}$.
Numerically we find that $\partial_t \varphi$ is very close to zero. The
maximum value is
\begin{equation}	
\max \partial_t \varphi = \max \frac{\partial_t \phi}{\phi} = 5\times 10^{-3} .
\end{equation}
This means that $\phi$ evolves on a timescale 
$T_{phi} = 2\pi (\phi /\partial_t \phi) > 1200 M^{ADM} $,
which is much longer than the dynamical timescale given by 
$T_{dyn} \sim 60 M^{ADM} $.
\begin{figure}
\epsfxsize=8.5cm 
\epsfbox{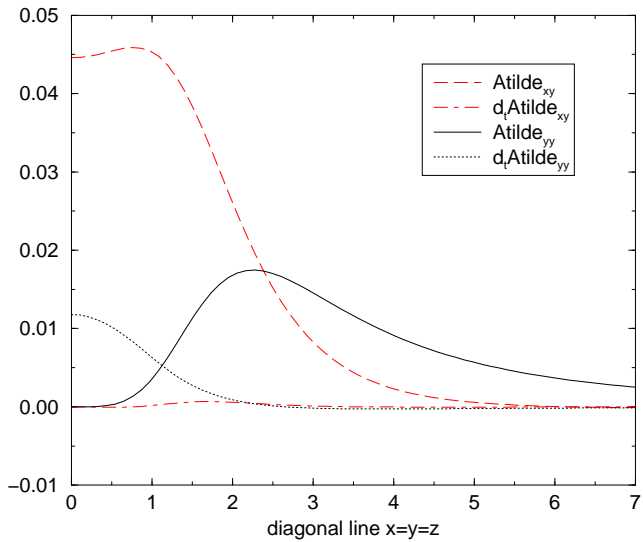}
\vspace{.4cm}
\caption{Selected components of $\Schlange{A}_{ij}$ and their time derivatives
along the diagonal $x=y=z$ for Pre-ISCO data. 
}
\label{Axygraph}
\end{figure}
\begin{figure}
\epsfxsize=8.5cm
\epsfbox{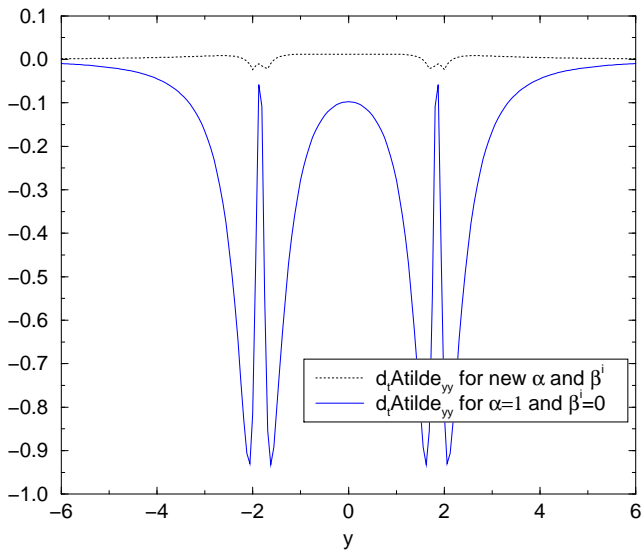}
\vspace{.4cm}
\caption{The time derivative of $\Schlange{A}_{yy}$ for ${\alpha}$ and
$\beta^i$ computed from Eqs.~(\ref{maxLapse}) and (\ref{ellipt_beta})
is much smaller than for ${\alpha}=1$ and $\beta^i =0$ for Pre-ISCO data.
}
\label{dtAyygraph}
\end{figure}

Finally, Fig.~\ref{Axygraph} shows the BSSN variable
$\Schlange{A}_{ij} = e^{-4\varphi} \bar{A}_{ij} = \phi^{-4}
\bar{A}_{ij}$, and its time derivative. We see that $\partial_t
\Schlange{A}_{ij}$ does not vanish. In fact for some components such
as $\Schlange{A}_{yy}$, $\partial_t \Schlange{A}_{ij}$ is not much
smaller then $\Schlange{A}_{ij}$ itself, while the time derivative of
the dominant component $\Schlange{A}_{xy}$ is small.  This implies
that puncture data cannot have a helical Killing vector.  However as
we can see in Fig.~\ref{dtAyygraph}, $\partial_t \Schlange{A}_{ij}$
is reduced by our choice of lapse and shift when compared to the case
of ${\alpha}=1$ and $\beta^i =0$, which is often used as initial
data for lapse and shift when evolving with dynamical gauge
conditions. Hence our choice of lapse and shift at least brings us
closer to the situation of minimal evolution.

\section{Discussion}
\label{Discussion}

We have shown that puncture data for quasicircular binary BHs obtained in
the CTT framework with the effective potential method allows a special gauge
choice that realizes some of the necessary conditions for the existence of
an approximate helical Killing vector field. Introducing a free parameter
for the lapse at the punctures we can satisfy the condition that the Komar
and ADM mass agree at spatial infinity. Since there are no further free
parameters in our gauge choice, it becomes a non-trivial fact that the Komar
and ADM mass also approximately agree at the punctures, and that the angular
velocity given by three different methods agree within a certain accuracy.
The angular velocity $\Omega_{\beta}$ determined by the shift that vanishes
at the punctures is equal to the angular velocity $\Omega$ inferred by
assuming that the Komar integral can be evaluated on any surface containing
the punctures, and both $\Omega_{\beta}$ and $\Omega$ agree approximately
with the angular velocity prescribed by the effective potential method.

Our gauge choice controls some of the time derivatives of the metric
variables, $\partial_t K = 0$ and $\partial_t\partial_j g^{ij} =
0$. Furthermore, we find that $ \phi $ varies on a timescale much
larger than the orbital timescale and that $g_{ij} $ varies on a
timescale which is at least five times the orbital timescale. The
magnitude of $ \partial_t \bar{A}_{ij} $ is reduced when we compare to
the case of lapse ${\alpha}=1$ and zero shift, but it still varies
on the orbital timescale.

One conclusion to draw from our results is that quasicircular puncture data
is not as far from possessing a helical Killing vector as one might
have expected. This is an important observation since to date all
gravitational wave forms obtained for binary BH mergers are
based on such initial data.

Note that in the present work we have used parameters for
quasicircular puncture data found in the literature. It should now be
possible to construct sequences of quasicircular orbits based on the
Komar and ADM mass equalities and compare with the effective potential
method. At a given separation $D$, one has to vary the linear momenta
$P^i_A$ and the constants $c_A$ for the lapse at the punctures until
both Eqs.~(\ref{MK_MADM}) and (\ref{MK_MADM1}) are satisfied. The
variation to find quasicircular orbits has to keep, for example, the
ADM masses at the punctures constant (cmp.\ \cite{Baker02a}), which
would be particularly natural in our approach.

Another direction for future work is to perform evolutions of
puncture data with the initial gauge derived here. Note that this is
not entirely straight-forward because the lapse crosses from positive
to negative values near the apparent horizon, which is also true for
the Meudon data sets construct in the CTS framework, but in that case
an isometry condition is imposed. It is not clear whether there are
numerical problems related to a negative lapse, even if one applies BH
excision techniques, but this is certainly worth exploring.

Finally, an immediate question is whether one can perhaps modify the
extrinsic curvature of the puncture data to better control $\partial_t
\bar{A}_{ij}$. Puncture data provides a
genuine technical simplification over methods with excision, but so
far it has not been possible to construct thin sandwich puncture data
in order to improve the extrinsic curvature. One aspect of the problem
becomes already apparent when considering maximal slicing of
Schwarzschild, which is known analytically~\cite{Estabrook73}. 
It is not hard to see that demanding $\partial_t \bar{g}_{ij}=0$ for all $t$
immediately leads to the standard isotropic coordinates with negative
lapse at the puncture and vanishing shift~\cite{Reimann03}.
The conclusion is that there is no maximal slicing of
Schwarzschild with static metric components and positive lapse. On the
other hand, in~\cite{Alcubierre02a} evolutions of Schwarzschild with
positive lapse have been performed that display an approximate Killing
vector, therefore the existence of thin sandwich puncture data has not
been ruled out.

\begin{acknowledgments}
It is a pleasure to thank Abhay Ashtekar and Greg Cook for
discussions. We acknowledge the support of the Center for
Gravitational Wave Physics funded by the National Science Foundation
under Cooperative Agreement PHY-01-14375. This work was also supported
by NSF grants PHY-02-18750 and PHY-98-00973.
\end{acknowledgments}

\appendix

\section{Integral $I_K$ defined without a Killing vector}
\label{I_K_withouKV}

If the conformal factor is chosen such that 
$\partial_t \mbox{det} g_{ij} = 0$,
the evolution equations for $\phi$, $g_{ij}$ and $K$ are
\begin{eqnarray}
\label{phi_evolution}
6 \partial_t \log \phi &=& -{\alpha} K + \bar{\nabla}_i \beta^i ,
\\
\label{g_evolution}
\partial_t g_{ij} &=& \phi^{-4}(\bar{L}\beta_{ij} - 2{\alpha}\bar{A}_{ij})
\\
\label{K_evolution}
\partial_t K &=& \bar{\nabla}^2 {\alpha}
		+ {\alpha}\bar{A}_{ij}\bar{A}^{ij}
		+ {\alpha}K^2 + \beta^i \partial_i K .
\end{eqnarray}    
If we solve Eqs.~(\ref{phi_evolution}) and
(\ref{g_evolution}) for ${\alpha} K$ and $\bar{A}_{ij}$ respectively
and insert the results
into Eq.~(\ref{K_evolution}) we obtain
\begin{eqnarray}  
\bar{\nabla}^i \left(\bar{\nabla}_i {\alpha} - \bar{K}_{ij} \beta^j \right)
&=& -\bar{A}^{ij}\phi^{4}\partial_t g_{ij} - \partial_t K \nonumber \\
&-& \left(6\partial_t \log \phi 
	-\frac{2}{3} \bar{\nabla}_i \beta^i \right) K ,
\end{eqnarray}                   
where we have made use of the momentum constraint.
Now note that the term in parentheses on the left hand side is the same
as in Eq.~(\ref{Komar2}). This suggests to define the volume integral
\begin{eqnarray}  
\label{I_V}
I_V&=& \frac{1}{4\pi} \int_V \bar{\nabla}^i \left(\bar{\nabla}_i {\alpha} 
			- \bar{K}_{ij} \beta^j \right) \nonumber \\
&=&  \frac{1}{4\pi} \int_V \Bigg[-\bar{A}^{ij}\phi^{4}\partial_t g_{ij} 
				 - \partial_t K        \nonumber \\
&& \ \ \ \   - \left(6\partial_t \log \phi 
            -\frac{2}{3} \bar{\nabla}_i \beta^i \right) K \Bigg] .
\end{eqnarray}
Next, let us define $I_K({\alpha}n+\beta,S)$ by the right hand side of
Eq.~(\ref{Komar2}) even if no Killing vector $\xi$ exists.
Now, if the volume
$V$ in Eq.~(\ref{I_V}) is bounded by the surfaces $S_{\infty}$ and 
$S_p$ we find that
\begin{equation}
\label{I_K_noKV}
I_K({\alpha}n+\beta,S_{\infty}) - I_K({\alpha}n+\beta,S_p) = I_V , 
\end{equation}
i.e.\ the value of $I_K({\alpha}n+\beta,S)$ in general does depend on the
surface $S$ on which it is evaluated. If, however,  
${\alpha}n^a+\beta^a$ is a Killing vector,
$I_V$ vanishes and $I_K({\alpha}n+\beta,S)$ does not depend on $S$. 

We now adjust the shift such that $\beta^i =0$ on $S_p$
and $\beta^i = \Omega_{\beta} \Phi^i $ on $S_{\infty}$, then
Eq.~(\ref{I_K_noKV}) becomes
\begin{equation}
I_K({\alpha}n,S_{\infty}) -2\Omega_{\beta} J^{ADM}_{\infty}
- I_K({\alpha}n ,S_p) = I_V .
\end{equation}
Choosing the lapse such that Eq.~(\ref{MK_MADM}) is satisfied,
yields
\begin{equation}
\label{Omega_FromIK}
\Omega_{\beta} = \frac{M^{ADM}_{\infty} 
- I_K(\xi,S_p)-I_V}{2J^{ADM}_{\infty}} .
\end{equation}
Hence $\Omega_{\beta}$ and $\Omega$ of Eq.~(\ref{OmegaFromI_K})
differ by the term $I_V$, which vanishes if ${\alpha}n^a+\beta^a$ is a
Killing vector. Another necessary condition for
the existence of a Killing vector is thus $\Omega_{\beta} = \Omega $.
For punctures with our choice of lapse we have  $K=\partial_t K = 0$,
so that all terms involving $K$ in $I_V$ of Eq.~(\ref{I_V}) vanish, 
the term with $\partial_t g_{ij}$ in $I_V$ however may not integrate to
zero, so that $\Omega_{\beta} = \Omega $ is non-trivial for puncture data.

\bibliography{references}

\begin{thebibliography}{32}
\expandafter\ifx\csname natexlab\endcsname\relax\def\natexlab#1{#1}\fi
\expandafter\ifx\csname bibnamefont\endcsname\relax
  \def\bibnamefont#1{#1}\fi
\expandafter\ifx\csname bibfnamefont\endcsname\relax
  \def\bibfnamefont#1{#1}\fi
\expandafter\ifx\csname citenamefont\endcsname\relax
  \def\citenamefont#1{#1}\fi
\expandafter\ifx\csname url\endcsname\relax
  \def\url#1{\texttt{#1}}\fi
\expandafter\ifx\csname urlprefix\endcsname\relax\def\urlprefix{URL }\fi
\providecommand{\bibinfo}[2]{#2}
\providecommand{\eprint}[2][]{\url{#2}}

\bibitem[{\citenamefont{Schutz}(1999)}]{Schutz99}
\bibinfo{author}{\bibfnamefont{B.}~\bibnamefont{Schutz}},
  \bibinfo{journal}{Class. Quantum Grav.} \textbf{\bibinfo{volume}{16}},
  \bibinfo{pages}{A131} (\bibinfo{year}{1999}).

\bibitem[{\citenamefont{Cook}(2000)}]{Cook:2000}
\bibinfo{author}{\bibfnamefont{G.~B.} \bibnamefont{Cook}},
  \bibinfo{journal}{Living Reviews in Relativity}
  \textbf{\bibinfo{volume}{2000}}, \bibinfo{pages}{5} (\bibinfo{year}{2000}).

\bibitem[{\citenamefont{Tichy et~al.}(2003)\citenamefont{Tichy, Br\"ugmann,
  Campanelli, and Diener}}]{Tichy02}
\bibinfo{author}{\bibfnamefont{W.}~\bibnamefont{Tichy}},
  \bibinfo{author}{\bibfnamefont{B.}~\bibnamefont{Br\"ugmann}},
  \bibinfo{author}{\bibfnamefont{M.}~\bibnamefont{Campanelli}},
  \bibnamefont{and} \bibinfo{author}{\bibfnamefont{P.}~\bibnamefont{Diener}},
  \bibinfo{journal}{Phys. Rev. D} \textbf{\bibinfo{volume}{67}},
  \bibinfo{pages}{064008} (\bibinfo{year}{2003}),
  \bibinfo{note}{gr-qc/0207011}.

\bibitem[{\citenamefont{Wilson and Mathews}(1995)}]{Wilson95}
\bibinfo{author}{\bibfnamefont{J.~R.} \bibnamefont{Wilson}} \bibnamefont{and}
  \bibinfo{author}{\bibfnamefont{G.~J.} \bibnamefont{Mathews}},
  \bibinfo{journal}{Phys. Rev. Lett.} \textbf{\bibinfo{volume}{75}},
  \bibinfo{pages}{4161} (\bibinfo{year}{1995}).

\bibitem[{\citenamefont{York}(1999)}]{York99}
\bibinfo{author}{\bibfnamefont{J.~W.} \bibnamefont{York}},
  \bibinfo{journal}{Phys. Rev. Lett.} \textbf{\bibinfo{volume}{82}},
  \bibinfo{pages}{1350} (\bibinfo{year}{1999}).

\bibitem[{\citenamefont{Gourgoulhon et~al.}(2001)\citenamefont{Gourgoulhon,
  Grandclement, and Bonazzola}}]{Gourgoulhon:2001ec}
\bibinfo{author}{\bibfnamefont{E.}~\bibnamefont{Gourgoulhon}},
  \bibinfo{author}{\bibfnamefont{P.}~\bibnamefont{Grandclement}},
  \bibnamefont{and}
  \bibinfo{author}{\bibfnamefont{S.}~\bibnamefont{Bonazzola}},
  \bibinfo{journal}{Phys. Rev. D} \textbf{\bibinfo{volume}{65}},
  \bibinfo{pages}{044020} (\bibinfo{year}{2001}), \eprint{gr-qc/0106015}.

\bibitem[{\citenamefont{Grandclement et~al.}(2001)\citenamefont{Grandclement,
  Gourgoulhon, and Bonazzola}}]{Grandclement:2001ed}
\bibinfo{author}{\bibfnamefont{P.}~\bibnamefont{Grandclement}},
  \bibinfo{author}{\bibfnamefont{E.}~\bibnamefont{Gourgoulhon}},
  \bibnamefont{and}
  \bibinfo{author}{\bibfnamefont{S.}~\bibnamefont{Bonazzola}},
  \bibinfo{journal}{Phys. Rev. D} \textbf{\bibinfo{volume}{65}},
  \bibinfo{pages}{044021} (\bibinfo{year}{2001}), \eprint{gr-qc/0106016}.

\bibitem[{\citenamefont{Pfeiffer et~al.}(2002)\citenamefont{Pfeiffer, Cook, and
  Teukolsky}}]{Pfeiffer:2002xz}
\bibinfo{author}{\bibfnamefont{H.~P.} \bibnamefont{Pfeiffer}},
  \bibinfo{author}{\bibfnamefont{G.~B.} \bibnamefont{Cook}}, \bibnamefont{and}
  \bibinfo{author}{\bibfnamefont{S.~A.} \bibnamefont{Teukolsky}},
  \bibinfo{journal}{Phys. Rev.} \textbf{\bibinfo{volume}{D66}},
  \bibinfo{pages}{024047} (\bibinfo{year}{2002}), \eprint{gr-qc/0203085}.

\bibitem[{\citenamefont{Cook}(2002)}]{Cook:2001wi}
\bibinfo{author}{\bibfnamefont{G.~B.} \bibnamefont{Cook}},
  \bibinfo{journal}{Phys. Rev.} \textbf{\bibinfo{volume}{D65}},
  \bibinfo{pages}{084003} (\bibinfo{year}{2002}),
  \eprint[http://arXiv.org/abs]{gr-qc/0108076}.

\bibitem[{\citenamefont{Brandt and Br{\"u}gmann}(1997)}]{Brandt97b}
\bibinfo{author}{\bibfnamefont{S.}~\bibnamefont{Brandt}} \bibnamefont{and}
  \bibinfo{author}{\bibfnamefont{B.}~\bibnamefont{Br{\"u}gmann}},
  \bibinfo{journal}{Phys. Rev. Lett.} \textbf{\bibinfo{volume}{78}},
  \bibinfo{pages}{3606} (\bibinfo{year}{1997}).

\bibitem[{\citenamefont{York}(1973)}]{York73}
\bibinfo{author}{\bibfnamefont{J.~W.} \bibnamefont{York}}, \bibinfo{journal}{J.
  Math. Phys.} \textbf{\bibinfo{volume}{14}}, \bibinfo{pages}{456}
  (\bibinfo{year}{1973}).

\bibitem[{\citenamefont{Cook}(1994)}]{Cook94}
\bibinfo{author}{\bibfnamefont{G.~B.} \bibnamefont{Cook}},
  \bibinfo{journal}{Phys. Rev. D} \textbf{\bibinfo{volume}{50}},
  \bibinfo{pages}{5025} (\bibinfo{year}{1994}).

\bibitem[{\citenamefont{Skoge and Baumgarte}(2002)}]{Skoge02}
\bibinfo{author}{\bibfnamefont{M.}~\bibnamefont{Skoge}} \bibnamefont{and}
  \bibinfo{author}{\bibfnamefont{T.}~\bibnamefont{Baumgarte}},
  \bibinfo{journal}{Phys. Rev. D} \textbf{\bibinfo{volume}{66}},
  \bibinfo{pages}{107501} (\bibinfo{year}{2002}).

\bibitem[{\citenamefont{Baker et~al.}(2002{\natexlab{a}})\citenamefont{Baker,
  Campanelli, Lousto, and Takahashi}}]{Baker:2002qf}
\bibinfo{author}{\bibfnamefont{J.}~\bibnamefont{Baker}},
  \bibinfo{author}{\bibfnamefont{M.}~\bibnamefont{Campanelli}},
  \bibinfo{author}{\bibfnamefont{C.~O.} \bibnamefont{Lousto}},
  \bibnamefont{and}
  \bibinfo{author}{\bibfnamefont{R.}~\bibnamefont{Takahashi}},
  \bibinfo{journal}{Phys. Rev.} \textbf{\bibinfo{volume}{D65}},
  \bibinfo{pages}{124012} (\bibinfo{year}{2002}{\natexlab{a}}),
  \eprint[http://arXiv.org/abs]{astro-ph/0202469}.

\bibitem[{\citenamefont{Laguna}(2003)}]{Laguna03}
\bibinfo{author}{\bibfnamefont{P.}~\bibnamefont{Laguna}}
  (\bibinfo{year}{2003}), \bibinfo{note}{in preparation}.

\bibitem[{\citenamefont{Alcubierre et~al.}(2001)\citenamefont{Alcubierre,
  Benger, Br\"ugmann, Lanfermann, Nerger, Seidel, and
  Takahashi}}]{Alcubierre00b}
\bibinfo{author}{\bibfnamefont{M.}~\bibnamefont{Alcubierre}},
  \bibinfo{author}{\bibfnamefont{W.}~\bibnamefont{Benger}},
  \bibinfo{author}{\bibfnamefont{B.}~\bibnamefont{Br\"ugmann}},
  \bibinfo{author}{\bibfnamefont{G.}~\bibnamefont{Lanfermann}},
  \bibinfo{author}{\bibfnamefont{L.}~\bibnamefont{Nerger}},
  \bibinfo{author}{\bibfnamefont{E.}~\bibnamefont{Seidel}}, \bibnamefont{and}
  \bibinfo{author}{\bibfnamefont{R.}~\bibnamefont{Takahashi}},
  \bibinfo{journal}{Phys. Rev. Lett.} \textbf{\bibinfo{volume}{87}},
  \bibinfo{pages}{271103} (\bibinfo{year}{2001}), \eprint{gr-qc/0012079}.

\bibitem[{\citenamefont{Alcubierre et~al.}(2003)\citenamefont{Alcubierre,
  Br\"ugmann, Diener, Koppitz, Pollney, Seidel, and Takahashi}}]{Alcubierre02a}
\bibinfo{author}{\bibfnamefont{M.}~\bibnamefont{Alcubierre}},
  \bibinfo{author}{\bibfnamefont{B.}~\bibnamefont{Br\"ugmann}},
  \bibinfo{author}{\bibfnamefont{P.}~\bibnamefont{Diener}},
  \bibinfo{author}{\bibfnamefont{M.}~\bibnamefont{Koppitz}},
  \bibinfo{author}{\bibfnamefont{D.}~\bibnamefont{Pollney}},
  \bibinfo{author}{\bibfnamefont{E.}~\bibnamefont{Seidel}}, \bibnamefont{and}
  \bibinfo{author}{\bibfnamefont{R.}~\bibnamefont{Takahashi}},
  \bibinfo{journal}{Phys. Rev. D} \textbf{\bibinfo{volume}{67}},
  \bibinfo{pages}{084023} (\bibinfo{year}{2003}), \eprint{gr-qc/0206072}.

\bibitem[{\citenamefont{Baker et~al.}(2002{\natexlab{b}})\citenamefont{Baker,
  Campanelli, and Lousto}}]{Baker:2001sf}
\bibinfo{author}{\bibfnamefont{J.}~\bibnamefont{Baker}},
  \bibinfo{author}{\bibfnamefont{M.}~\bibnamefont{Campanelli}},
  \bibnamefont{and} \bibinfo{author}{\bibfnamefont{C.~O.}
  \bibnamefont{Lousto}}, \bibinfo{journal}{Phys. Rev.}
  \textbf{\bibinfo{volume}{D65}}, \bibinfo{pages}{044001}
  (\bibinfo{year}{2002}{\natexlab{b}}),
  \eprint[http://arXiv.org/abs]{gr-qc/0104063}.

\bibitem[{\citenamefont{Baker et~al.}(2001)\citenamefont{Baker, Br{\"u}gmann,
  Campanelli, Lousto, and Takahashi}}]{Baker:2001nu}
\bibinfo{author}{\bibfnamefont{J.}~\bibnamefont{Baker}},
  \bibinfo{author}{\bibfnamefont{B.}~\bibnamefont{Br{\"u}gmann}},
  \bibinfo{author}{\bibfnamefont{M.}~\bibnamefont{Campanelli}},
  \bibinfo{author}{\bibfnamefont{C.~O.} \bibnamefont{Lousto}},
  \bibnamefont{and}
  \bibinfo{author}{\bibfnamefont{R.}~\bibnamefont{Takahashi}},
  \bibinfo{journal}{Phys. Rev. Lett.} \textbf{\bibinfo{volume}{87}},
  \bibinfo{pages}{121103} (\bibinfo{year}{2001}),
  \eprint[http://arXiv.org/abs]{gr-qc/0102037}.

\bibitem[{\citenamefont{Baker et~al.}(2003)\citenamefont{Baker, Campanelli,
  Lousto, and Takahashi}}]{Baker:2003ds}
\bibinfo{author}{\bibfnamefont{J.}~\bibnamefont{Baker}},
  \bibinfo{author}{\bibfnamefont{M.}~\bibnamefont{Campanelli}},
  \bibinfo{author}{\bibfnamefont{C.~O.} \bibnamefont{Lousto}},
  \bibnamefont{and} \bibinfo{author}{\bibfnamefont{R.}~\bibnamefont{Takahashi}}
  (\bibinfo{year}{2003}), \eprint[http://arXiv.org/abs]{astro-ph/0305287}.

\bibitem[{\citenamefont{Shibata and Nakamura}(1995)}]{Shibata95}
\bibinfo{author}{\bibfnamefont{M.}~\bibnamefont{Shibata}} \bibnamefont{and}
  \bibinfo{author}{\bibfnamefont{T.}~\bibnamefont{Nakamura}},
  \bibinfo{journal}{Phys. Rev. D} \textbf{\bibinfo{volume}{52}},
  \bibinfo{pages}{5428} (\bibinfo{year}{1995}).

\bibitem[{\citenamefont{Baumgarte and Shapiro}(1999)}]{Baumgarte99}
\bibinfo{author}{\bibfnamefont{T.~W.} \bibnamefont{Baumgarte}}
  \bibnamefont{and} \bibinfo{author}{\bibfnamefont{S.~L.}
  \bibnamefont{Shapiro}}, \bibinfo{journal}{Physical Review D}
  \textbf{\bibinfo{volume}{59}}, \bibinfo{pages}{024007}
  (\bibinfo{year}{1999}), \eprint{gr-qc/9810065}.

\bibitem[{\citenamefont{Baumgarte}(2000)}]{Baumgarte00a}
\bibinfo{author}{\bibfnamefont{T.~W.} \bibnamefont{Baumgarte}},
  \bibinfo{journal}{Phys. Rev. D} \textbf{\bibinfo{volume}{62}},
  \bibinfo{pages}{024018} (\bibinfo{year}{2000}), \eprint{gr-qc/0004050}.

\bibitem[{\citenamefont{Baker et~al.}(2000)\citenamefont{Baker, Br\"ugmann,
  Campanelli, and Lousto}}]{Baker00b}
\bibinfo{author}{\bibfnamefont{J.}~\bibnamefont{Baker}},
  \bibinfo{author}{\bibfnamefont{B.}~\bibnamefont{Br\"ugmann}},
  \bibinfo{author}{\bibfnamefont{M.}~\bibnamefont{Campanelli}},
  \bibnamefont{and} \bibinfo{author}{\bibfnamefont{C.~O.}
  \bibnamefont{Lousto}}, \bibinfo{journal}{Class. Quantum Grav.}
  \textbf{\bibinfo{volume}{17}}, \bibinfo{pages}{L149} (\bibinfo{year}{2000}).

\bibitem[{\citenamefont{Br{\"u}gmann}(2000)}]{Bruegmann99b}
\bibinfo{author}{\bibfnamefont{B.}~\bibnamefont{Br{\"u}gmann}},
  \bibinfo{journal}{Ann. Phys. (Leipzig)} \textbf{\bibinfo{volume}{9}},
  \bibinfo{pages}{227} (\bibinfo{year}{2000}), \bibinfo{note}{gr-qc/9912009}.

\bibitem[{\citenamefont{Komar}(1959)}]{Komar59}
\bibinfo{author}{\bibfnamefont{A.}~\bibnamefont{Komar}},
  \bibinfo{journal}{Phys. Rev.} \textbf{\bibinfo{volume}{113}},
  \bibinfo{pages}{934} (\bibinfo{year}{1959}).

\bibitem[{\citenamefont{Ashtekar and Magnon-Ashtekar}(1979)}]{Ashtekar79a}
\bibinfo{author}{\bibfnamefont{A.}~\bibnamefont{Ashtekar}} \bibnamefont{and}
  \bibinfo{author}{\bibfnamefont{A.}~\bibnamefont{Magnon-Ashtekar}},
  \bibinfo{journal}{J. Math. Phys.} \textbf{\bibinfo{volume}{20}},
  \bibinfo{pages}{793} (\bibinfo{year}{1979}).

\bibitem[{\citenamefont{Wald}(1984)}]{Wald84}
\bibinfo{author}{\bibfnamefont{R.~M.} \bibnamefont{Wald}},
  \emph{\bibinfo{title}{General Relativity}} (\bibinfo{publisher}{The
  University of Chicago Press}, \bibinfo{address}{Chicago},
  \bibinfo{year}{1984}).

\bibitem[{\citenamefont{Alcubierre and Br\"ugmann}(2001)}]{Alcubierre00a}
\bibinfo{author}{\bibfnamefont{M.}~\bibnamefont{Alcubierre}} \bibnamefont{and}
  \bibinfo{author}{\bibfnamefont{B.}~\bibnamefont{Br\"ugmann}},
  \bibinfo{journal}{Phys. Rev. D} \textbf{\bibinfo{volume}{63}},
  \bibinfo{pages}{104006} (\bibinfo{year}{2001}), \eprint{gr-qc/0008067}.

\bibitem[{\citenamefont{Baker}(2002)}]{Baker02a}
\bibinfo{author}{\bibfnamefont{B.~D.} \bibnamefont{Baker}}
  (\bibinfo{year}{2002}), \eprint[http://arXiv.org/abs]{gr-qc/0205082}.

\bibitem[{\citenamefont{Estabrook et~al.}(1973)\citenamefont{Estabrook,
  Wahlquist, Christensen, DeWitt, Smarr, and Tsiang}}]{Estabrook73}
\bibinfo{author}{\bibfnamefont{F.}~\bibnamefont{Estabrook}},
  \bibinfo{author}{\bibfnamefont{H.}~\bibnamefont{Wahlquist}},
  \bibinfo{author}{\bibfnamefont{S.}~\bibnamefont{Christensen}},
  \bibinfo{author}{\bibfnamefont{B.}~\bibnamefont{DeWitt}},
  \bibinfo{author}{\bibfnamefont{L.}~\bibnamefont{Smarr}}, \bibnamefont{and}
  \bibinfo{author}{\bibfnamefont{E.}~\bibnamefont{Tsiang}},
  \bibinfo{journal}{Phys. Rev. D} \textbf{\bibinfo{volume}{7}},
  \bibinfo{pages}{2814} (\bibinfo{year}{1973}).

\bibitem[{\citenamefont{Reimann}(2003)}]{Reimann03}
\bibinfo{author}{\bibfnamefont{B.}~\bibnamefont{Reimann}}, Master's thesis,
  \bibinfo{school}{Universit\"at Potsdam} (\bibinfo{year}{2003}).

\end{thebibliography}

\end{document}